\documentclass[manuscript]{aastex}

\shorttitle{Analysis of global coronal
disturbances} \shortauthors{Muhr et al.}

\begin{document}


\title{Analysis of characteristic parameters of large-scale coronal waves
observed by STEREO/EUVI}


\author{N. Muhr\altaffilmark{1}, A.M. Veronig\altaffilmark{1}, I.W. Kienreich\altaffilmark{1}, M. Temmer\altaffilmark{1}}
\affil{IGAM (Institute of Geophysics, Astrophysics and Meteorology), Institute of Physics, University of Graz,
       \it Universit\"atsplatz 5, A--8010 Graz, Austria}

\and

\author{B. Vr\v{s}nak\altaffilmark{2}}
\affil{Hvar Observatory, Faculty of Geodesy, University of Zagreb,
       \it Ka\v{c}i\'{c}eva 26, HR--10000 Zagreb, Croatia}
       \email{nicole.muhr@uni-graz.at}




\begin{abstract}
The kinematical evolution of four EUV waves, well observed by the Extreme
UltraViolet Imager (EUVI) onboard the Solar-Terrestrial Relations Observatory (STEREO), is studied by visually tracking
the wave fronts as well
as by a semi-automatized perturbation profile method leading to results matching each other
within the error limits.
The derived mean velocities of the events under study
lie in the range of 220--350~km\,s$^{-1}$. The fastest of the events (May 19, 2007) reveals a significant deceleration of $\approx -190$~m~s$^{-2}$ while the others are consistent with a constant velocity during the wave propagation.
The evolution of the maximum intensity values reveals initial intensification by 20 up to 70\%, and decays to original levels within 40--60~min, while the width at half maximum and full
maximum of the perturbation profiles are broadening by a factor of 2\,--\,4. The integral below the perturbation profile remains basically constant in two cases,
while it shows a decrease by a factor of 3\,--\,4 in the other two cases. From the peak perturbation amplitudes we estimate the
corresponding magneto-sonic Mach numbers M$_{ms}$ which are in the range of 1.08--1.21.
The perturbation profiles reveal three distinct features behind the propagating wave fronts: coronal dimmings, stationary brightenings and rarefaction regions. All of them appear after the wave passage and are only slowly fading away. Our findings indicate that the events under study are weak shock fast-mode MHD waves initiated by the CME lateral expansion.
\end{abstract}


\section{Introduction}
Large-scale, large-amplitude disturbances propagating through the solar corona have
been first observed by \citet{moses97} and \citet{thompson98} in
images recorded by the Extreme Ultraviolet Imaging Telescope
\citep[EIT;][]{delaboudiniere95} onboard the Solar and Heliospheric
Observatory \citep[SoHO;][]{domingo95}, thereafter called `EIT waves'. They were
originally interpreted as the coronal counterparts of the
chromospheric Moreton waves \citep{moretonramsey60} as suggested by
\citet{uchida68} with his coronal fast-mode MHD wave interpretation of Moreton waves.
Though this interpretation has been confirmed for several case studies
\citep[e.g.][]{thompson99, warmuth01, pohjolainen01, vrsnak02, khan02, vrsnak06, veronig06, muhr10},
the statistical characteristics of
EUV and Moreton waves are quite different. In particular, EIT waves
generally propagate at much lower velocities,
$v\approx200$--400~km\,s$^{-1}$, than Moreton waves,
$v\approx1000$~km\,s$^{-1}$ \citep[e.g.][]{klassen00, biesecker02, thompson09}. Thus, there
is still an ongoing debate of whether EIT waves are really the
coronal counterparts of Moreton waves. Additionally, it is still an open discussion
whether they are caused by the flare explosive energy release or the
erupting CME \citep[e.g.][]{warmuth01, warmuth04b, zhukov04, cliver05, vrsnak08} and whether they are waves
at all or
rather propagating disturbances related to magnetic field line
opening and restructuring associated with the CME liftoff
\citep[e.g.][]{delanee99, chen02, attrill07, wills07}.

Numerical simulations of EIT waves as fast-mode
magnetohydrodynamical (MHD) waves resulted in
wave phenomena mimicking observational data quite closely
\citep{wang00, wu01, ofman02, ofman07}. An interesting by-product of
the simulations by \cite{ofman02} and \cite{terradas04} are stationary
brightenings in the vicinity of the launch sites, in line with
observations of such areas at the fronts of EIT waves \citep{delanee99, delanee00, attrill07}.
\citet{delannee07} suggested them to be a
result of magnetic field restructuring during the CME liftoff
generated either by Joule heating or an increase in density due to
plasma compression. Simulations by \citet{cohen09} favored the latter interpretation
with some additional effect due to increased temperature resulting
from plasma compression.

Until the launch of the STEREO \citep[Solar-Terrestrial
Relations Observatory,][]{kaiser08} mission in 2006, with the Extreme
UltraViolet Imager \citep[EUVI;][]{howard08} onboard, observations of EIT
waves were drastically limited by a $\approx12$~min cadence of the
EIT instrument in the 195~$\hbox{\AA}$ passband. The identically built
instruments EUVI-A and EUVI-B onboard the twin STEREO spacecraft
observe the solar corona in four different EUV passbands with a high
observing cadence (up to 75~s) and a large field-of-view (up to
1.7~R$_{\hbox{$\odot$}}$) from two different vantage points.

Studies of large-scale waves with STEREO/EUVI reveal velocities in
the range of $\approx200$--400~km\,s$^{-1}$ and a decelerating
character consistent with freely propagating large-amplitude MHD
fast-mode waves \citep{veronig08, veronig10, long08, kienreich09}.
With simultaneous observations by the two STEREO spacecraft, it was
for the first time possible to analyze the 3D nature of EIT waves
with stereoscopic techniques \citep{kienreich09, patsourakos09,
patsourakos09b, ma09, temmer11}. \citet{veronig08},
\citet{long08} and \citet{gopalswamy09} reported the refraction and
reflection of a wave at the border of a coronal hole providing
strong evidence for the wave nature of the phenomenon. However,
\citet{attrill10} questioned these findings, favoring instead the hypothesis of two
simultaneously launched waves from two distinctively separated
sites. In the EUVI wave event of 2010 January 17, studied by
\citet{veronig10}, first observations of the
full 3D wave dome and its lateral and radial expansion could be performed.
In a recent study that became available during the revision process of the present paper, \citet{long11} analyze the kinematical aspects and the evolution of the full width at half maximum (FWHM) and integral of the 2007 May 19 and 2009 February 13 events with similar methods. Therefore we will discuss and compare their findings with our results in section~\ref{sec:discussion}.
For recent
reviews regarding the kinematics of large-scale EUV waves, their
morphology and relationship to associated solar phenomena we refer
to \citet{wills10} and \citet{gallagher10}.

In this paper, we study high-cadence EUV observations of four well
defined EUV waves observed by STEREO/EUVI that occurred between May
2007 and April 2010. These waves are special due to their pronounced
amplitudes although they occurred during the extreme current solar
minimum.
Our study has the following main aims:
\begin{enumerate}
\item We apply and compare two different methods to derive the wave kinematics: visual tracking of the foremost part of wave front and a method based on the intensity profiles of the propagating perturbation.
\item Based on the perturbation profiles, we derive quantities which provide insight into the physical character of the phenomenon in terms of Mach numbers as well as evolution of the amplitude, width and integrated intensity of the disturbances.
\item The perturbation profiles are further studied with respect to associated phenomena, such as stationary
brightenings, coronal dimmings and rarefaction
regions behind the wave fronts.
\end{enumerate}

\section{Data and Observations}
The four large-scale coronal waves under study were recorded by EUVI on 2007 May 19, 2009 February 13, 2010 January 17, and 2010 April 29. The EUVI instruments are part of the Sun Earth
Connection Coronal and Heliospheric Investigation
\citep[SECCHI;][]{howard08} instrument suite on board the STEREO-A
(Ahead; ST-A) and STEREO-B (Behind; ST-B) spacecraft.
The separation angle between ST-A and ST-B steadily increases by
$\approx$45$\degr$ per year. EUVI observes the chromosphere and
low corona in four spectral passbands (He~II 304~$\hbox{\AA}$: T$\sim0.07$~MK;
Fe~IX 171~$\hbox{\AA}$: T$\sim1$~MK; Fe~XII
195~$\hbox{\AA}$: T$\sim1.5$~MK; Fe~XV 284~$\hbox{\AA}$: T$\sim2.25$~MK)
out to 1.7~R$_{\hbox{$\odot$}}$ with a pixel-limited spatial
resolution of 1$\arcsec$.6 pixel$^{-1}$ \citep{wuelser04}.
For the kinematical analysis we used all passbands if possible, while for the
perturbation profiles only the 195~$\hbox{\AA}$ images were studied, where the wave signal
is most prominent.

The EUVI wave event of 2007 May 19 was recorded by both STEREO
spacecrafts (ST-A and ST-B)
and was observable from the Earth view.
High-cadence EUVI images in the 171~$\hbox{\AA}$ (75~s), 195~$\hbox{\AA}$ (600~sec),
284~$\hbox{\AA}$ (300~s) and 304~$\hbox{\AA}$ (300~sec) channels are available. This event was the
first distinct EUV wave observed by EUVI, and was already studied in quite some detail by several authors
\citep{long08, veronig08, gopalswamy09, kerdraon10, long11} revealing a propagation velocity decelerating from
$\approx450$ to 200~km\,s$^{-1}$ as well as an associated type II burst indicating shock
formation in the corona.

The event of 2009 February 13 was observed by both STEREO
spacecraft, and occurred in perfect quadrature (separation angle of
$\approx90\degr$). The wave was observed on disk in ST-B and on
the limb in ST-A. These observations provided a unique basis to
study the 3D nature of the wave. \cite{patsourakos09} and
\cite{kienreich09} determined a propagation height of the wave of $\approx
100$~Mm, with a mean (on-disk) velocity of
$\approx230$~km\,s$^{-1}$. For the analysis, 171~$\hbox{\AA}$ and
195~$\hbox{\AA}$ filtergrams of ST-B are available with a cadence of
300~s and 600~s, respectively.

On 2010 January 17, ST-A and ST-B were 134$\degr$ apart from each
other. The wave was observed in the eastern hemisphere of ST-B, which was
situated 70$\degr$ behind Earth on its orbit around the Sun. This
event was unique due to the full 3D wave dome structure that was
observed in the different EUVI channels with a lateral (on-disk)
propagation velocity of $\approx300$~km\,s$^{-1}$ and an upward propagation
velocity of $\approx600$~km\,s$^{-1}$ \citep{veronig10}. The EUVI-B
imaging cadence is 2.5 minutes in the 171~$\hbox{\AA}$, 5 minutes in
the 195~$\hbox{\AA}$, 2.5$-–$5 minutes in the 284~$\hbox{\AA}$, and 5
minutes in the 304~$\hbox{\AA}$ passband.

The event of 2010 April 29 was the last and strongest of four
homologous waves observed by ST-B within a period of 8 hours
at a position angle of 70$\degr$
behind solar limb as seen from Earth \citep{kienreich11}. EUVI images in all four wavelengths are
available during the event. However, the cadence of two hours in the
171~$\hbox{\AA}$ and 284~$\hbox{\AA}$ channels limited a possible wave
tracking to the 195~$\hbox{\AA}$ and 304~$\hbox{\AA}$ channels. Due to
a lack of wave signatures in the 304~$\hbox{\AA}$ channel, only
195~$\hbox{\AA}$ filtergrams with a cadence of 300~s are used for the
analysis.

All EUVI filtergrams were reduced using the SECCHI$\_$PREP routines
available within SolarSoft. Furthermore, we differentially rotated
each data set to a common reference time. In order to enhance faint
coronal wave signatures, we derived running ratio (RR) images
dividing each image by a frame taken 10 minutes earlier, as
well as base-ratio (BR) images dividing each image by the
last pre-event image. Finally, a median filter was applied to the
images to remove small scale variations. RR images are used for the
visual tracking of the coronal waves, since in RR images the signal
of propagating disturbances is highest. The perturbation profiles
are calculated from BR images, which provide better insight
into the changes of physical parameters due to the passing wave front.

\begin{figure}[h]
    \centering%
    \includegraphics[width=0.45\textwidth,keepaspectratio=true]{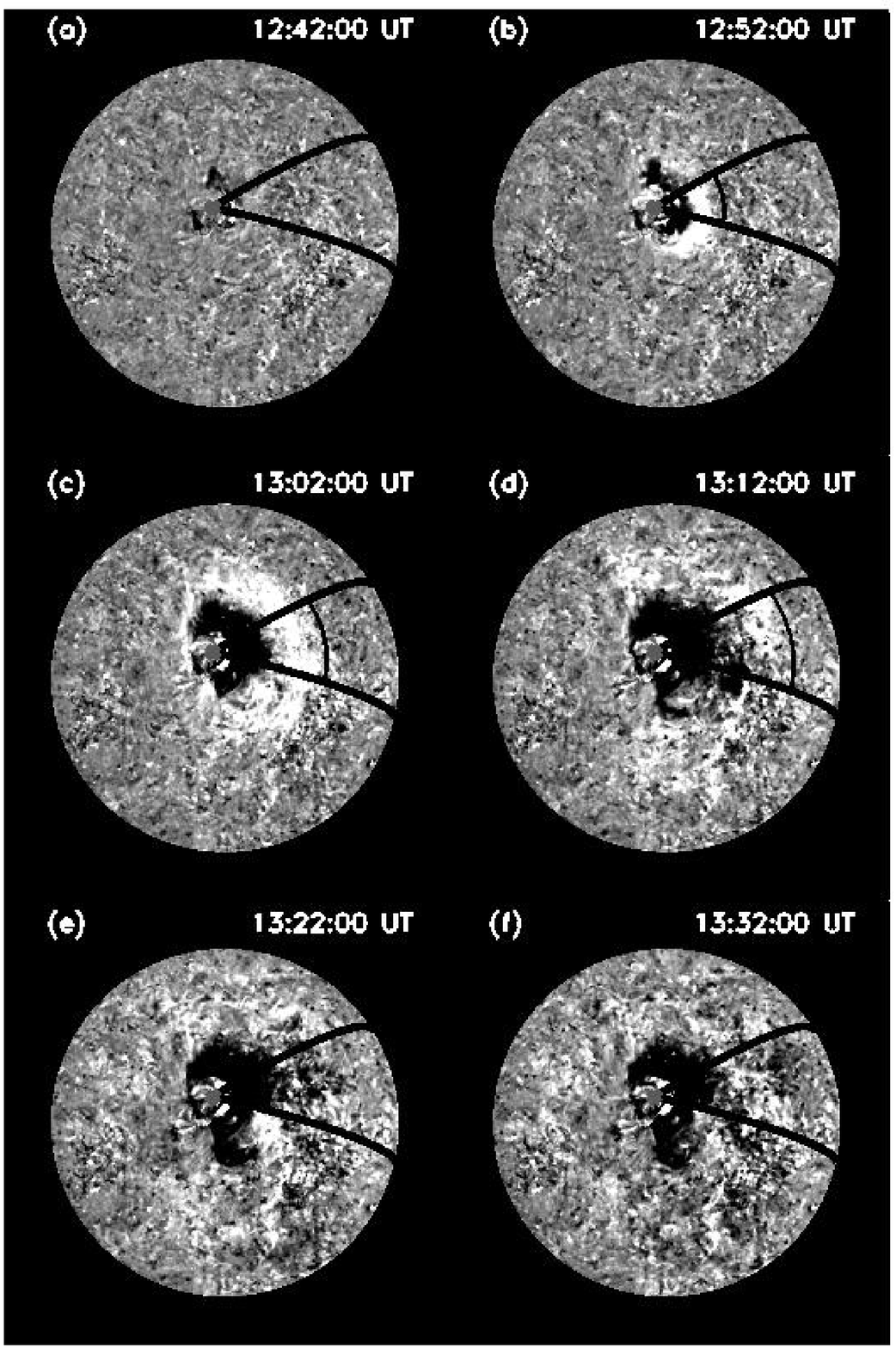}
    \caption[.]{Sequence of STEREO/EUVI-A BR images of the 2007 May 19 event. The determined initiation center for the EUVI wave
    is indicated by the cross. The black lines indicate great circles through the wave center determining the propagation direction of $10\degr\pm22.5\degr$, on which we focus the wave analysis. The black fronts correspond to the positions of the
    leading edges of the wave fronts extracted from the perturbation profiles. The
    field of view (FoV) is x=[$-$1200$\arcsec$, +1200$\arcsec$], y=[$-$1200$\arcsec$, +1200$\arcsec$] with
    the origin at the center of the Sun.} \label{img:20070519195Abaseratio_rot0_overview}
\end{figure}

\begin{figure}[h]
    \centering%
    \includegraphics[width=0.45\textwidth,keepaspectratio=true]{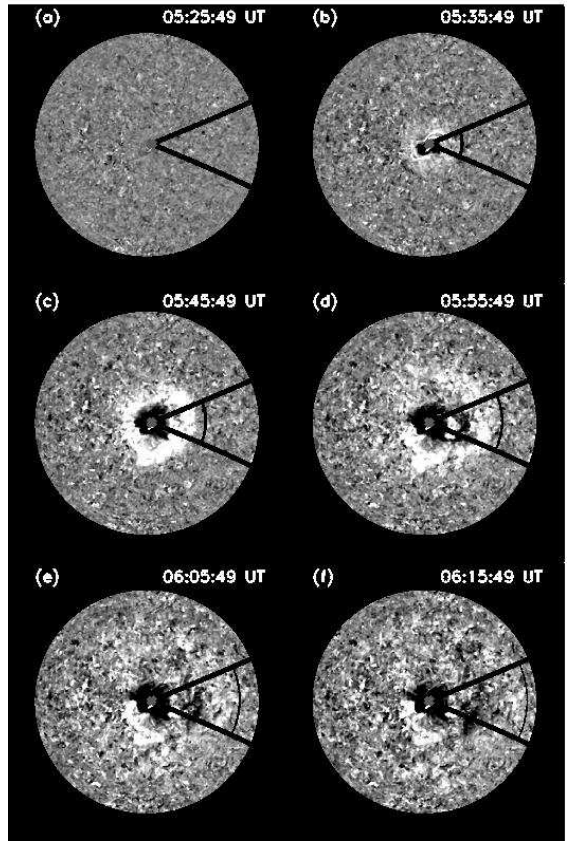}
    \caption[.]{Same as Figure~\ref{img:20070519195Abaseratio_rot0_overview} but for the 2009 February 13 event, with the propagation direction of $340\degr\pm22.5\degr$.} \label{img:20090213195Bbaseratio_rot0_overview}
\end{figure}

\begin{figure}[h]
    \centering%
    \includegraphics[width=0.45\textwidth,keepaspectratio=true]{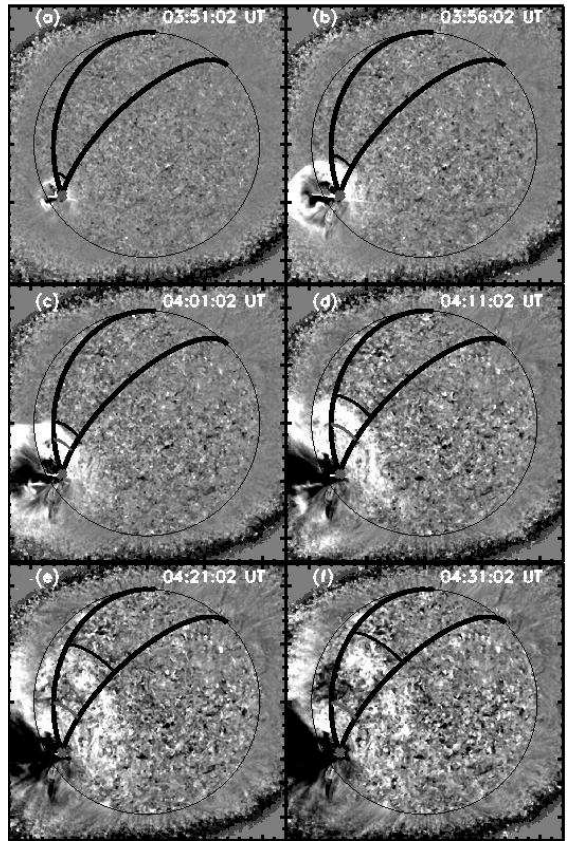}
    \caption[.]{Same as Figure~\ref{img:20100117195Bbaseratio_rot0_overview} but for the 2010 January 17 event, with the propagation direction of $65\degr\pm22.5\degr$.} \label{img:20100117195Bbaseratio_rot0_overview}
\end{figure}

\begin{figure}[h]
    \centering%
    \includegraphics[width=0.45\textwidth,keepaspectratio=true]{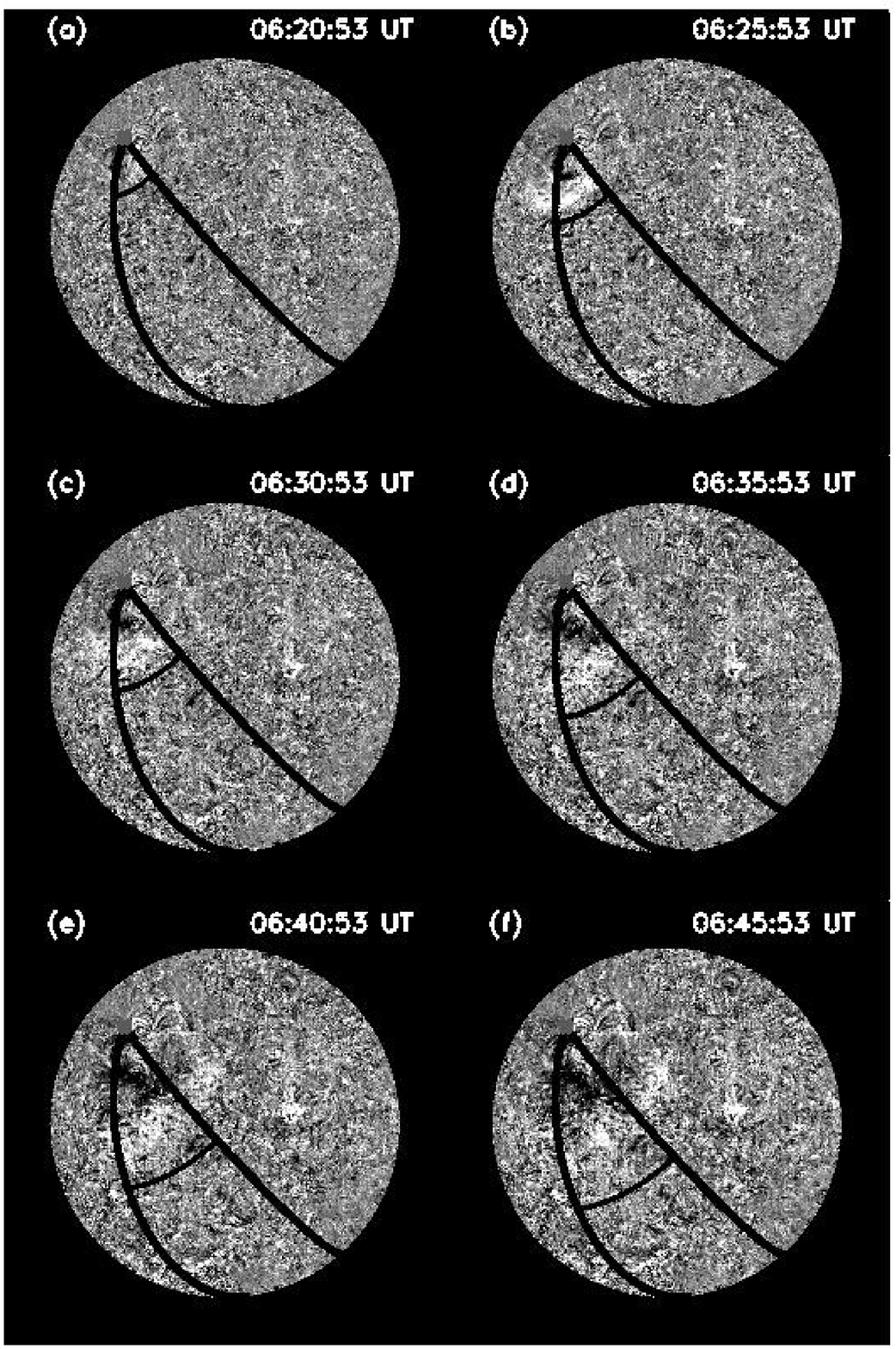}
    \caption[.]{Same as Figure~\ref{img:20100428195Bbaseratio_3_rot0_overview} but for the 2010 April 29 event, with the propagation direction of $300\degr\pm22.5\degr$.} \label{img:20100428195Bbaseratio_3_rot0_overview}
\end{figure}

\section{Analysis}
The kinematical analysis of large-scale propagating disturbances is
usually based on visual tracking of the outer edge of the wave
front.
Thus, the results are severely influenced by the observer and his/her
interpretation of the wave front.
Therefore, more objective and reproducible as well as automated
methods are desirable. Two alternative
semi-automated methods have been applied so far to study EUV waves: the Huygens plotting method
\citep{wills99} and the perturbation profile method
\citep{warmuth04a, podladchikova05, muhr10, veronig10}.

In this study we focus on the perturbation profile
method because of several reasons. First, this semi-automated method provides
insight into important physical parameters of the EUVI wave (amplitude, widths, Mach
number). Second, it provides a semi-automated alternative to the reconstruction of the
wave kinematics by visual tracking. Third, detailed analysis
of the perturbation profiles provide information on associated
phenomena not observable by eye, which can give us important
insight into the underlying physics of the events. In the
following, we describe the two different methods, i.e. the visual tracking method
and perturbation profiles, and their application to the study of EUV waves.

\subsection{Kinematics via visual tracking}\label{sec:kin_via_vis}
We visually tracked the wave fronts in series of EUVI RR images. The
eruption center was derived by applying circular fits to the
earliest observed EUVI wave fronts on the 3D solar surface
appearing as ellipses in the 2D projected image
\citep{veronig06}. In order to enhance the statistical significance
and to ensure a realistic error estimate for the wave center
position, all available passbands were used to determine a mean value
and standard deviation. For each event, we focussed for both, the visual and the
perturbation profile method, on the same specific 45$\degr$ propagation
sector, in which the disturbance is most pronounced. The
chosen sectors are restricted by two great circles (parts of them
are shown by black curves in
Figures~\ref{img:20070519195Abaseratio_rot0_overview}--\ref{img:20100428195Bbaseratio_3_rot0_overview}), which pass through
the determined wave center. To obtain the wave kinematics, we calculated for each point of the wave
front within the selected propagation sector its distance from the
eruption center along great circles on the solar surface, and then
averaged over them.

\subsection{Kinematics by perturbation profiles}\label{sec:kin_via_pro}
The second approach to analyze the EUV wave propagation is based on the
perturbation profiles. The method starts at
the eruption center determined from the visually tracked wave fronts
(see section~\ref{sec:kin_via_vis}) and sums the intensity values of
all pixels between two constantly growing concentric circles (again
in the deprojected heliospheric plane) defining annuli with a radial
width of 1$\degr$ within the selected propagation sector, which span over an
angular width of 45$\degr$ (for an illustration see movie~1 of the
online version). Due to the fact that the area over which we sum up
the pixel values is steadily growing by moving to greater radii,
each annulus is averaged over its pixel sum. This is resulting in a
mean intensity as a function of distance measured along the solar
surface from the wave center. The procedure is repeated for each
frame, i.e. time step, until the EUV wave fades away.

In perturbation profiles, propagating disturbances can be
identified as a distinct bump above the background level (intensity
level of 1.0 in base ratio images). Modified Gaussian envelopes emphasizing the leading part of the
wave bump are then fitted to the perturbation profiles for two reasons. First of all,
perturbation profiles show roughly a Gaussian form
\citep{wills06, veronig10, kienreich11} and secondly it is
very useful for the robustness and automatization of the latter part of
the profile algorithm, in which the positions of the leading edges of the waves are extracted. \citet{downs11} state that the peak intensity may be strongly affected by the emission of CME material, whereas the foremost front corresponds to the actual wave. The trailing edge of the wave is also strongly influenced by features like stationary brightenings, flare emission, CME and dimming regions behind the wave front. Therefore we use modified Gaussian envelopes to the original data set. First of all, the original data is examined for the peak intensity. Then a simple Gaussian fit is applied to the first half of the wave bump from the maximum until the leading edge. Finally, the Gaussian envelope is mirrored at the position of the maximum. The result is a Gaussian fit matching the foremost leading edge of the wave fronts much better than simple Gaussian fits while for the rest of the wave bump a characteristic envelope shape is formed.

From the modified Gaussian fits to the perturbation profiles we derive the
position of the wave front. Starting at the maximum amplitude level of
each wave bump we define the leading edge of the wave fronts at the
position where the Gaussian envelope to the profile falls below an
intensity level of $I/I_{0}=1.02$, with $I$ the intensity entries of the
current image and $I_0$ the intensity entries of the base image.
This value is reasonable due to
the fact that the human eye is able to identify intensity changes of
as small as 1--2\% above the background level. Thus,
this should provide us with a good measure when comparing the wave front positions
from the perturbation profiles and the visual tracking.

\subsection{Calculation of perturbation profile parameters}
It is possible to gain useful information on the wave characteristics by extracting
distinct wave parameters from the perturbation profiles. We derived the amplitude, the FW,
the FWHM and the integral below the perturbation profile
(down to the intensity ratio of 1.02).
The FW of the wave is defined
as the width of the Gaussian envelope, whereas the FWHM is defined
as the distance between those two points where the intensity level drops
below 50\% of the maximum value. The integral of the perturbation
profiles is the area below the Gaussian fits.

The calculation of the magnetosonic Mach numbers is based on the peak
perturbation amplitude values, $A_{\rm{max}}$, of each wave event, derived from the peak of the Gaussian
fits to the profiles. Due to the definition of base ratio images,
we know that an increase in the perturbation amplitude above the
background level can be interpreted as $A=I/I_0$, where the intensity $I$ in an
optically thin coronal spectral line is given as
\begin{equation}
I=\int_h f(T, n_e) n_e^2dh\,,
\end{equation}
with the
emission measure $EM=\int_h n_e^2dh$ along the line of sight and the contribution function $f(T, n_e)$
depending on plasma temperature $T$ and electron density $n_e$ \citep{phillips08}.
To obtain an intensity value in a relative broad EUV passband, like the EUVI 195~$\hbox{\AA}$ passband,
the integral over several spectral lines has to be conducted. Since there is a temperature
and density dependence for each spectral line, the following approximations are made: First, the
integration is only along the pressure scale height $H$ with an average constant density $n$ \citep{wills06}. Thus the intensity turns into $I\approx f(T, n_e) n_e^2 H$, with $n_e^2 H$ the emission measure $EM$. Second,
the quiet Sun in the 195~$\hbox{\AA}$ passband is dominated by its strongest line at 195.12~$\hbox{\AA}$ \citep{zanna03}
which is formed in a narrow temperature range of 1.2~MK to 1.8~MK, with the peak response at $\approx1.4$~MK \citep{feldman99}. In Fig.~3 of \citet{zhukov11} it is evident that the contribution function of the 195.12~$\hbox{\AA}$ spectral line only weakly depends on density, i.e. $f(T, n_e)\approx f(T)$, and only weakly depends on temperature near its peak response which is also the position where the differential emission measure peaks. Thus and under the assumption that the temperature does not
change considerably by the passing wave front, the relative intensity change $I/I_0$ (with $I$ and $I_0$ referring to the initial and final state, respectively) can be used to estimate the density increase $n/n_0\sim (I/I_0)^{1/2}$. \citet{zhukov11} gives a detailed discussion of this approach under consideration of the usage of the CHIANTI atomic database.

For each wave we estimate the density jump
$X_c=n/n_0$ at the peak amplitude of the perturbation profile. Assuming the shock height to be in the low
solar corona \citep{veronig10}, we can conclude from typical EUV wave velocities in the range of 200$-$500~km\,s$^{-1}$
that the assumption of a quasi-perpendicular fast-mode MHD shock is
reasonable \citep{mann99}. Considering the Rankine-Hugoniot jump
conditions at a perpendicular shock front \citep[e.g.,][]{priest82},
the magnetosonic Mach number $M_{ms}=v_{c}/v_{ms}$ (where $v_{c}$ is
the coronal shock velocity and $v_{ms}$ the magnetosonic speed) can
be expressed as
\begin{equation}\label{equ:machnumber}
 M_{ms}= \sqrt{\frac{X_c (X_c +5+5\beta_c)}{(4-X_c)(2+5\beta_c/3)}}\,,
\end{equation}
where $\beta_c$ is the ambient coronal plasma-to-magnetic pressure
ratio, $X_c=n/n_0$ is the density jump at the shock front, and for the
specific-heat ratio (the polytropic index) we substituted $\gamma=5/3$. For
the plasma beta in the quiet Sun we assume a value of
$\beta_c=0.1$ and note that Eq.~\ref{equ:machnumber} depends only weakly on $\beta$ \citep{vrsnak02}.

\subsection{Coronal dimmings, stationary brightenings and rarefaction regions}
Beside the propagating bright coronal wave front, the perturbation profiles show
some other distinct features characterized either through persistent intensity
depression (coronal dimmings and rarefaction regions) or intensity enhancement
(stationary brightenings).

Coronal dimmings are regions of dramatically decreased plasma density occurring after
the CME lift-off. They are usually interpreted as plasma evacuation due to the field line
opening during the CME eruption \citep{hudson96, harra01, harra07}, and may last for several hours up to a day.
We stress that there are different types of coronal dimmings, those in the near vicinity of the
eruption site, referred to as core coronal dimmings, and so-called secondary dimmings further away from the eruption site \citep{mandrini07, muhr10}. In this paper we deal with core coronal dimmings. They usually consist of two areas marking the
footpoints of the erupting flux rope \citep{mandrini05, crooker06, muhr10} and show an intensity decrease of
$\approx40$--60~\% \citep[e.g.][]{chertok05}.

Stationary brightenings form at the edge of core coronal dimmmings
and are noticed as intensity enhancements \citep[e.g.][]{cohen09}. They
are formed after the wave passage and last for a period of
several tens of minutes until fading away. A possible explanation
for their formation is that they are a result of material flow,
storage and compression at this specific location due to an
obstacle that can not be overcome easily \citep{delanee99,
delannee07}.

Rarefaction regions are expected to occur right behind a propagating
wave pulse, clearly visible by an intensity amplitude below the
background intensity \citep{landau87}. The moving wave pulse is a
region of compressed plasma, continuously compressing the plasma in
front of itself. Consequently, in the section behind the wave pulse a rarefaction
region is formed, expanding to a certain distance. The wave travels
at a velocity higher than the surrounding Alfv$\acute{e}$n velocity
and thus the region behind the pulse is thinned out. The pressure
and density in this region falls below the equilibrium values. The
rarefaction region is the result of this evacuation due to the wave
propagation. In contrast to the coronal dimming, which may be
stationary and in most cases relatively huge, the rarefaction region
propagates following
the wave pulse and can
be noticed as a small, localized dip in the rear section of the
perturbation profile between the wave pulse and the coronal dimming.

\section{Results}
\subsection{Wave kinematics}
In
Figures~\ref{img:20070519195Abaseratio_rot0_overview}--\ref{img:20100428195Bbaseratio_3_rot0_overview},
the morphology and evolution of the four EUVI wave events under
study is shown in BR images. We note that the wave fronts are not
observed in the near vicinity of the eruption center but are first
visible at distances of $\approx240$~Mm, 90~Mm, 150~Mm and 260~Mm
from the respective initiation centers of 2007 May 19, 2009 February 13,
2010 January 17, 2010 April 29 event and can be followed up to distances of
600--900~Mm.

The top panels of
Figures~\ref{img:all20070519195Arunratio_1_rot0_kinematics_amp_fwh_int_4wave_gauss}--\ref{img:all20100428195Brunratio_3_rot0_kinematics_amp_fwh_int_4wave_gauss}
show the distance-time plots of the events. We display the
wave kinematics derived from both methods (visually tracking and perturbation profiles) together with error bars,
linear and quadratic least square fits to the visually tracked wavefronts, as well
as 95\% confidence intervals and prediction boundaries of the
linear fits. The mean velocities derived for the four events under study lie in the
range of 220--350~km\,s$^{-1}$.

Three of the four events propagate with constant speed. The small deceleration values derived from the quadratic fits are in the range of $-6$ to $-13$~m~s$^{-2}$, which corresponds to a deceleration of $\approx5\%$ with respect to the start velocity values derived from the quadratic fit to the wave kinematics. This lies within the confidence intervals, and thus it is justified to represent the kinematics of the three events by a propagation with constant velocity over the full propagation distance. We note that the kinematical analysis of the perturbation profiles of these three events reveal a somewhat stronger deceleration as compared to the visual tracking method, but they are still only marginally significant corresponding to a deceleration rate of $\approx10\%$ of the start velocity values.
Only for the 2007 May 19 event, we obtain a significant deceleration of $-193$~m~s$^{-2}$, which
corresponds to $30\%$ of the start velocity. This is also the fastest wave in our sample with a mean velocity of $\approx350$~km\,s$^{-1}$. In Table~\ref{tab:table1},
the derived information on the propagation characteristics (velocity, deceleration) are summarized.

\begin{table*}[ht]
\centering
\scriptsize{
\begin{tabular}{lcccccccccc}
  \hline\hline
     Event & S/C & \multicolumn{2}{c}{v$_{\rm{lin}}$ [km\,s$^{-1}$]}  & \multicolumn{2}{c}{v$_{\rm{quad,0}}$ [km\,s$^{-1}$]} & \multicolumn{2}{c}{$a_0$ [m~s$^{-2}$]} & $I_{\rm{max}}$ &$X_c$& $M_{\rm{ms}}$ \\
     && vis. & prof.& vis. & prof. & vis. & prof.&&&\\
     \hline
  2007 May 19 &ST-A& 348$\pm$29 & 338$\pm$18 & 483$\pm$73 & 429$\pm$58 & $-193.2\pm87$ & $-84.7\pm60$ & 1.61 &1.27& 1.21 \\
  2009 Feb 13 &ST-B& 228$\pm$17 & 221$\pm$36 & 237$\pm$72 & 261$\pm$72 & $-6.9\pm25$ & $-23.4\pm30$ & 1.39 &1.18& 1.14 \\
  2010 Jan 17 &ST-B& 286$\pm$7  & 275$\pm$17 & 300$\pm$24 & 314$\pm$53 & $-13.4\pm38$ & $-39.4\pm49$  & 1.59 &1.26& 1.20 \\
  2010 Apr 29 &ST-B& 337$\pm$27 & 321$\pm$23 & 342$\pm$88 & 361$\pm$76 & $-7.6\pm52$  & $-51.7\pm93$ & 1.22 &1.12& 1.08 \\
  \hline
\end{tabular}
\caption{EUVI wave properties of the events under study. We list the mean velocity values v$_{\rm{lin}}$ derived from linear fits, as well as the start velocity values v$_{\rm{quad,0}}$ and the acceleration values $a_0$ derived from the quadratic fits for both methods, i.e. visual tracking method and profile method. Additionally, the peak intensity-amplitudes
extracted from the perturbation profiles together with the density jump values $X_c$ and the calculated Mach numbers $M_{\rm{ms}}$ are listed.}}
\label{tab:table1}
\end{table*}

We note that the position measurements of both methods match each other
quite well, the maximum differences lie within an average of 20--40~Mm.
As a consequence, also the velocity values derived via visual tracking
correspond to these obtained by the perturbation profiles, and both values agree
within the error limits (see Table~\ref{tab:table1}).

\subsection{Overall characteristics of perturbation profiles}
The consistent kinematical results obtained with both methods suggest the
perturbation profile method to be an adequate alternative to
the visual tracking method. In particular, it is an additional possibility
to analyze large-scale disturbances in the solar atmosphere by
producing kinematical results not burdened by subjective judgements of the observer. Thus,
the perturbation-profile results are
easier to reproduce than those derived by the
visual-tracking method.

\begin{figure}
    \centering%
    \includegraphics[width=0.7\textwidth,keepaspectratio=true]{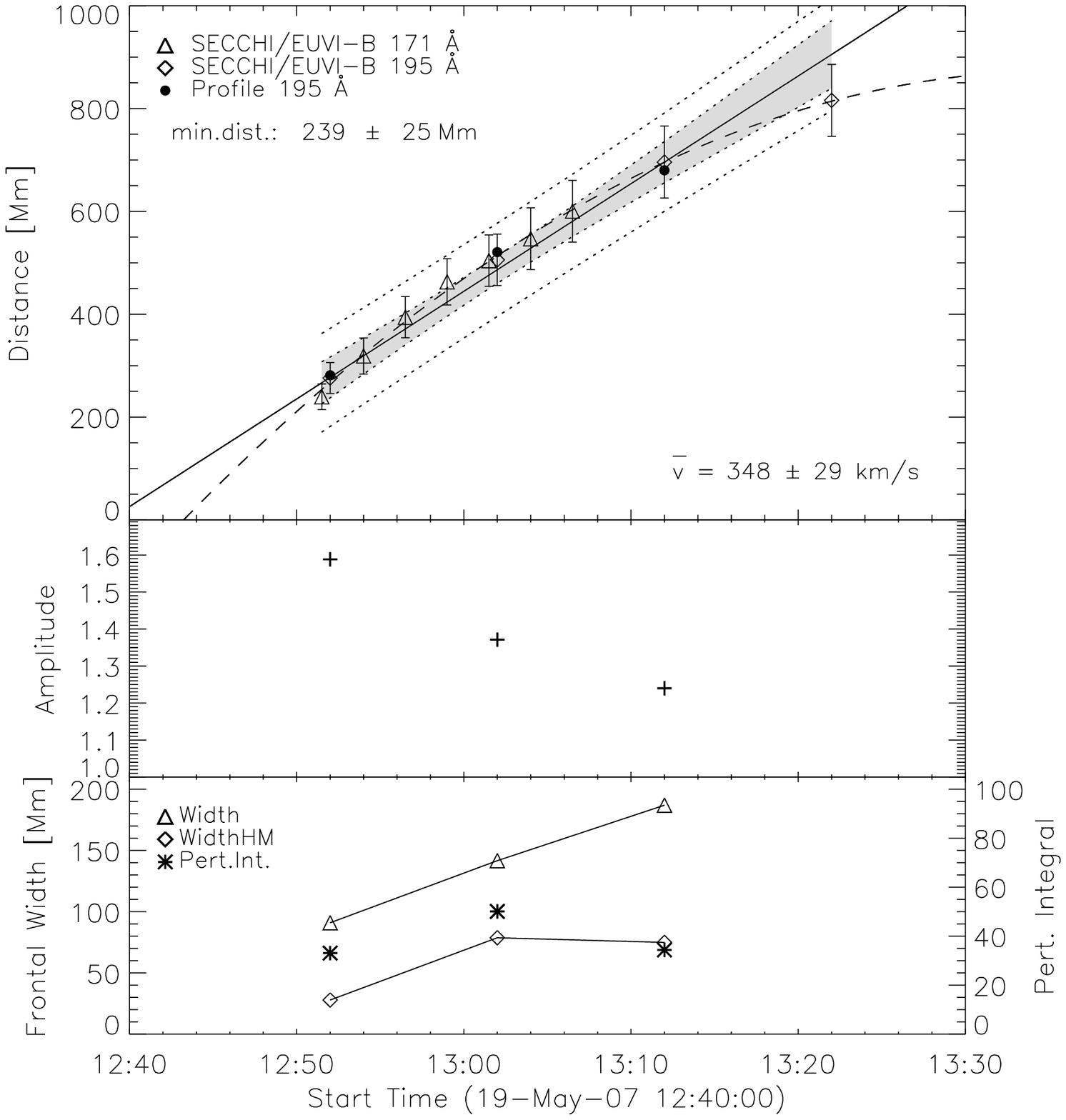}
    \caption[.]{Top: Kinematics of the wave fronts of the 2007 May 19 wave observed on the solar disk
    in the EUVI-A 171$\hbox{\AA}$ (triangles) and 195$\hbox{\AA}$ (diamonds) channels by the visual tracking method together with the kinematical measurements we derived from the perturbation profiles (full circles). The error bars added to the measurements represent the diffusiveness of the wave fronts. The solid and dashed lines indicate the linear and quadratic least square fits to the data set obtained by visual tracking, respectively.
    The dotted lines indicate the 95\% confidence interval (filled grey area) and the prediction boundary of the linear fit.
    Middle: Evolution of the amplitude of the perturbation profiles shown in
Figure~\ref{img:all20070516195Bprofile_1_rot0panel10}. Bottom: Evolution of the frontal width of the wave pulse (triangles),
the full width at half maximum (FWHM, squares) and the integral of the gaussian envelopes (asterisks).}
\label{img:all20070519195Arunratio_1_rot0_kinematics_amp_fwh_int_4wave_gauss}
\end{figure}

\begin{figure}
    \centering%
    \includegraphics[width=0.7\textwidth,keepaspectratio=true]{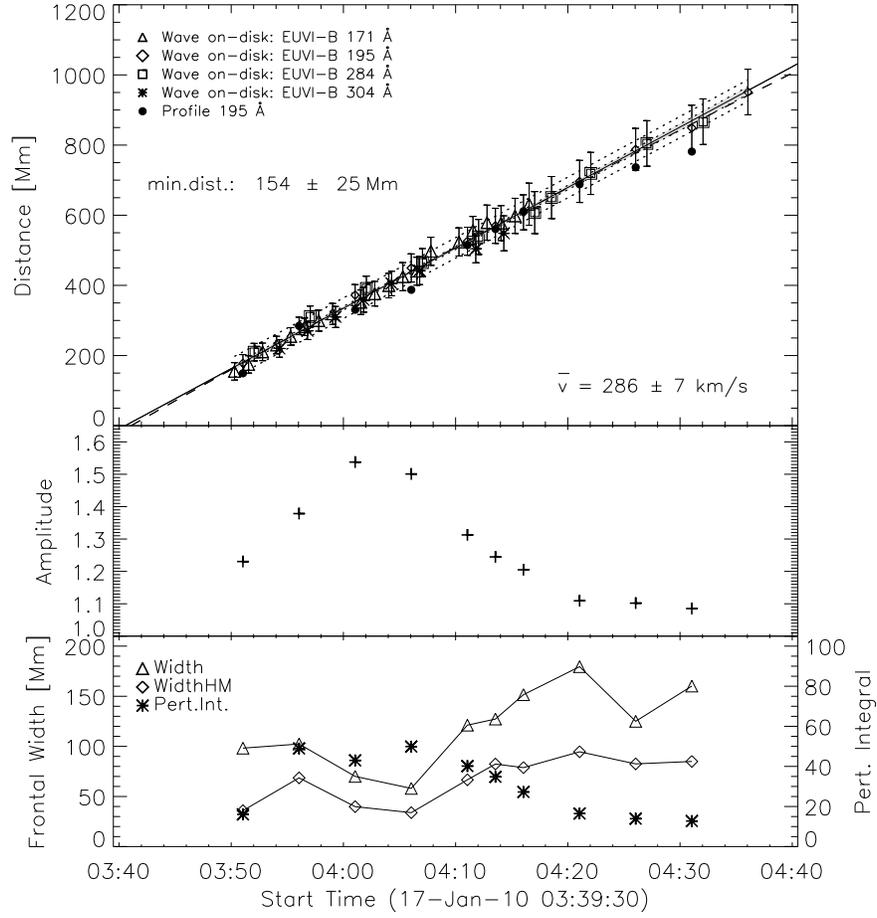}
    \caption[.]{Same as in Fig.~\ref{img:all20070519195Arunratio_1_rot0_kinematics_amp_fwh_int_4wave_gauss} but for the 2010 January 17 wave. Top: Kinematics of the wave fronts are derived from the EUVI-B 171$\hbox{\AA}$ (triangles), 195$\hbox{\AA}$ (diamonds), 284$\hbox{\AA}$ (squares) and 304$\hbox{\AA}$ (asterisks) observations.}
\label{img:all20100117195Brunratio_2_rot0_kinematics_amp_fwh_int_4wave_gauss}
\end{figure}

\begin{figure}
    \centering%
    \includegraphics[width=0.7\textwidth,keepaspectratio=true]{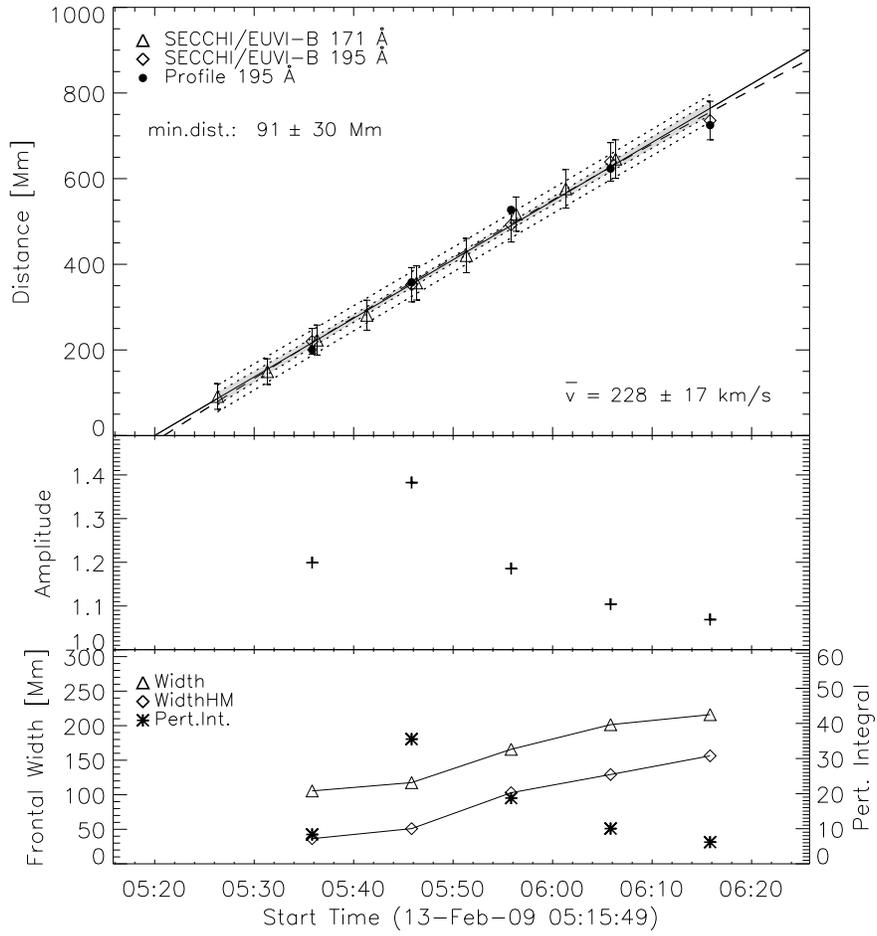}
    \caption[.]{Same as in Fig.~\ref{img:all20070519195Arunratio_1_rot0_kinematics_amp_fwh_int_4wave_gauss} but for the 2009 February 13 wave.}
\label{img:all20090213195Brunratio_1_rot0_kinematics_amp_fwh_int_4wave_gauss}
\end{figure}

\begin{figure}
    \centering%
    \includegraphics[width=0.7\textwidth,keepaspectratio=true]{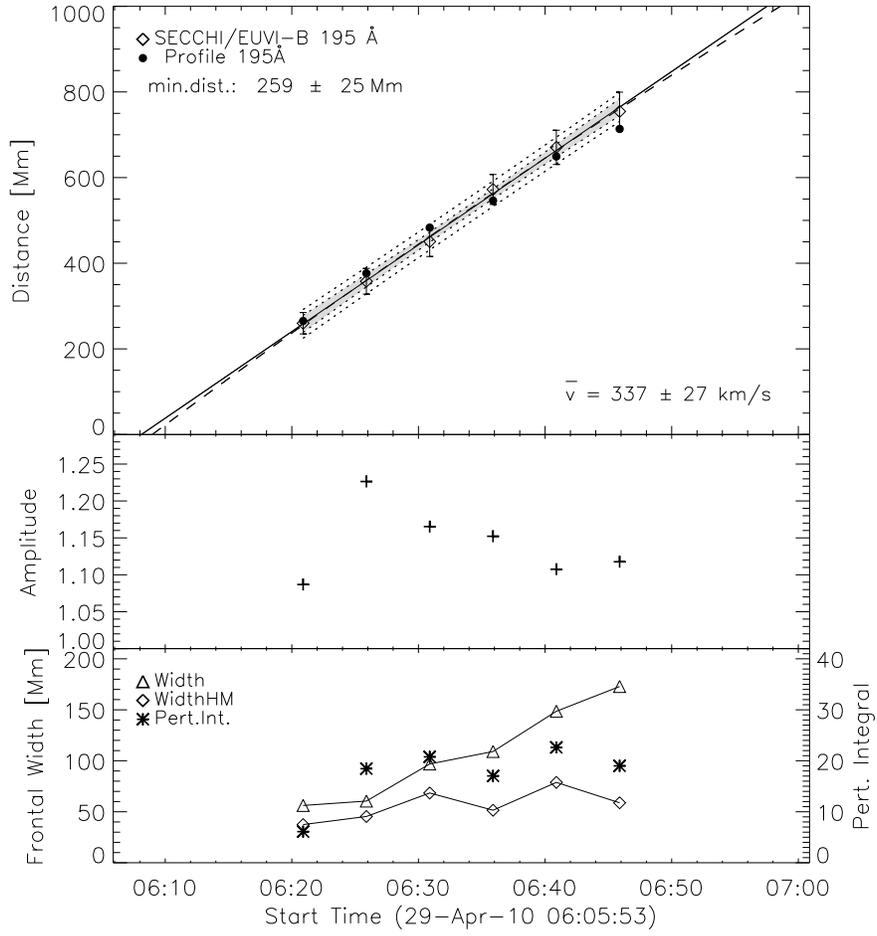}
    \caption[.]{Same as in Fig.~\ref{img:all20070519195Arunratio_1_rot0_kinematics_amp_fwh_int_4wave_gauss} but for the 2010 April 28 wave.}
\label{img:all20100428195Brunratio_3_rot0_kinematics_amp_fwh_int_4wave_gauss}
\end{figure}

\begin{figure}
    \centering%
    \includegraphics[width=0.3\textwidth,keepaspectratio=true]{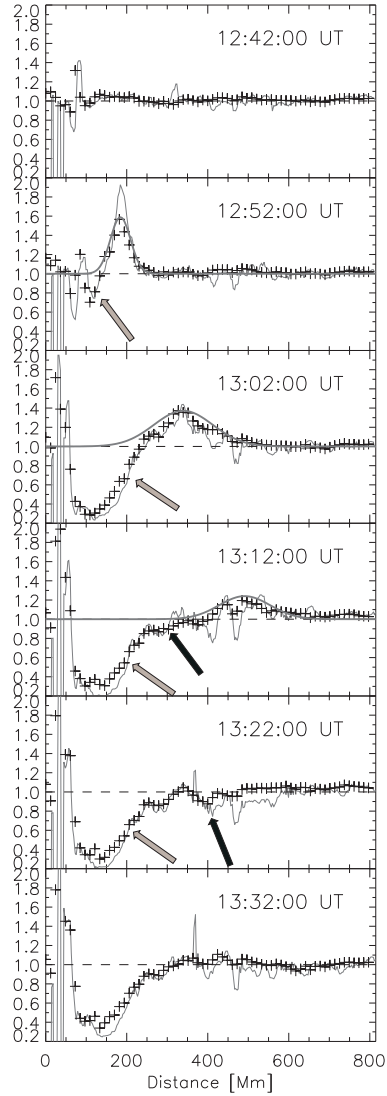}
    \caption[.]{Perturbation profiles of 2007 May 19 ST-A observations in the analyzed sector ($10\degr\pm22.5\degr$)
    together with the gaussian envelopes
    of the wave pulses (thick grey curves).
    The thin grey perturbation profiles correspond to the narrow sector of 8$\pm$2$\degr$.
    The grey arrows point to the core dimming while the black arrows point to the rarefaction region.} \label{img:all20070516195Bprofile_1_rot0panel10}
\end{figure}

Moreover, we can quantify various important wave parameters from
the perturbation profiles. Figures~\ref{img:all20070516195Bprofile_1_rot0panel10}\,--\,\ref{img:all20100428195Bprofile_3_rot0panel10} show for the four wave events under study
the evolution of the perturbation profiles. In all four cases
we observe a clear
intensification in their early
phase reaching its peak amplitude $A_{\rm{max}}$ value around
10\,--\,20 minutes (second or third panel
of Figures~\ref{img:all20070516195Bprofile_1_rot0panel10}\,--\,\ref{img:all20100428195Bprofile_3_rot0panel10})
after the first remarkable
wave bump can be observed. The evolution of the intensity amplitudes of all
events under study is displayed in
the middle panels of
Figures~\ref{img:all20070519195Arunratio_1_rot0_kinematics_amp_fwh_int_4wave_gauss}--\ref{img:all20100428195Brunratio_3_rot0_kinematics_amp_fwh_int_4wave_gauss}. The peak
perturbation amplitudes $A_{\rm{max}}$ lie
in the range of 1.2$-$1.6. The
calculated density jump values, $X_c$, are therefore in the range of
1.1$-$1.3, and the peak magnetosonic Mach numbers, $M_{ms}$,
lie in the range of 1.08$-$1.21 (Table~\ref{tab:table1}).

In addition, we derived the full width of the wave as well as the
FWHM. The evolution of the pulse width is plotted in
the bottom panel of
Figures~\ref{img:all20070519195Arunratio_1_rot0_kinematics_amp_fwh_int_4wave_gauss} --
\ref{img:all20100428195Brunratio_3_rot0_kinematics_amp_fwh_int_4wave_gauss}
for the four wave events under study. The width of the wave pulse as well
as the FWHM increases during its evolution by a factor of 2--3,
whereas the integral below the perturbation profile remains basically constant
for the events of 2007 May 19 and 2010 April 29, and decreases
for the events of 2009 February 13 and 2010 January 17 by a factor of 4 and 3, respectively.

\begin{figure}
    \centering%
    \includegraphics[width=0.3\textwidth,keepaspectratio=true]{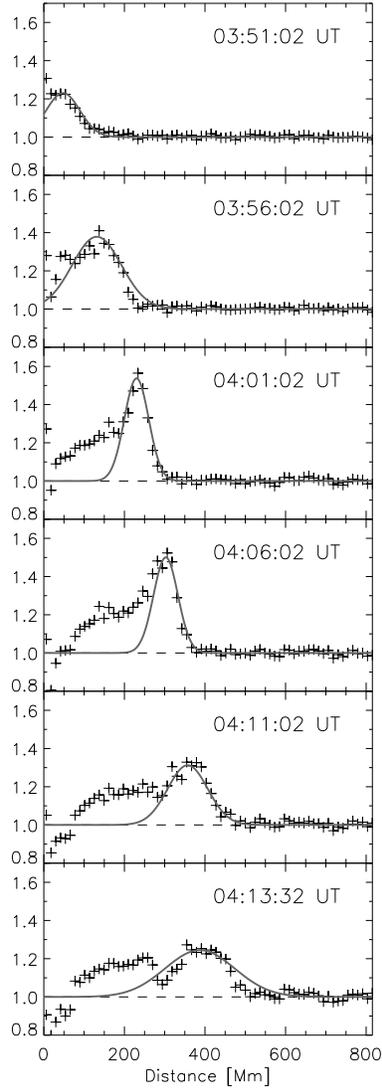}
    \caption[.]{Perturbation profiles of 2010 January 17 ST-B observations in the analyzed sector ($65\degr\pm22.5\degr$)
    together with the gaussian envelopes
    of the wave pulses (thick grey curves).}
\label{img:all20100117195Bprofile_2_rot0panel10}
\end{figure}

\begin{figure}
    \centering%
    \includegraphics[width=0.3\textwidth,keepaspectratio=true]{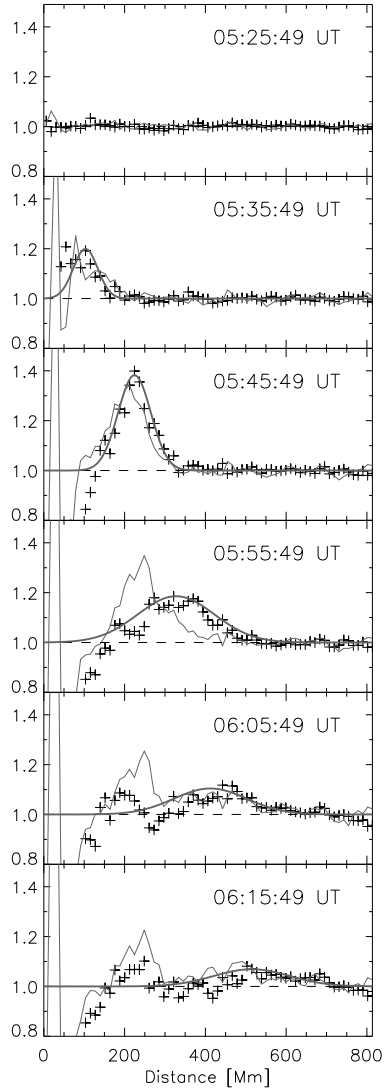}
    \caption[.]{Perturbation profiles of 2009 February 13 ST-A observations in the analyzed sector ($45\degr\pm22.5\degr$)
    together with the gaussian envelopes
    of the wave pulses (thick grey curves).
    The thin grey perturbation profiles correspond to the sector of $270\degr\pm22.5\degr$.}
\label{img:all20090213195Aprofile_1_rot0panel10}
\end{figure}

\begin{figure}
    \centering%
    \includegraphics[width=0.3\textwidth,keepaspectratio=true]{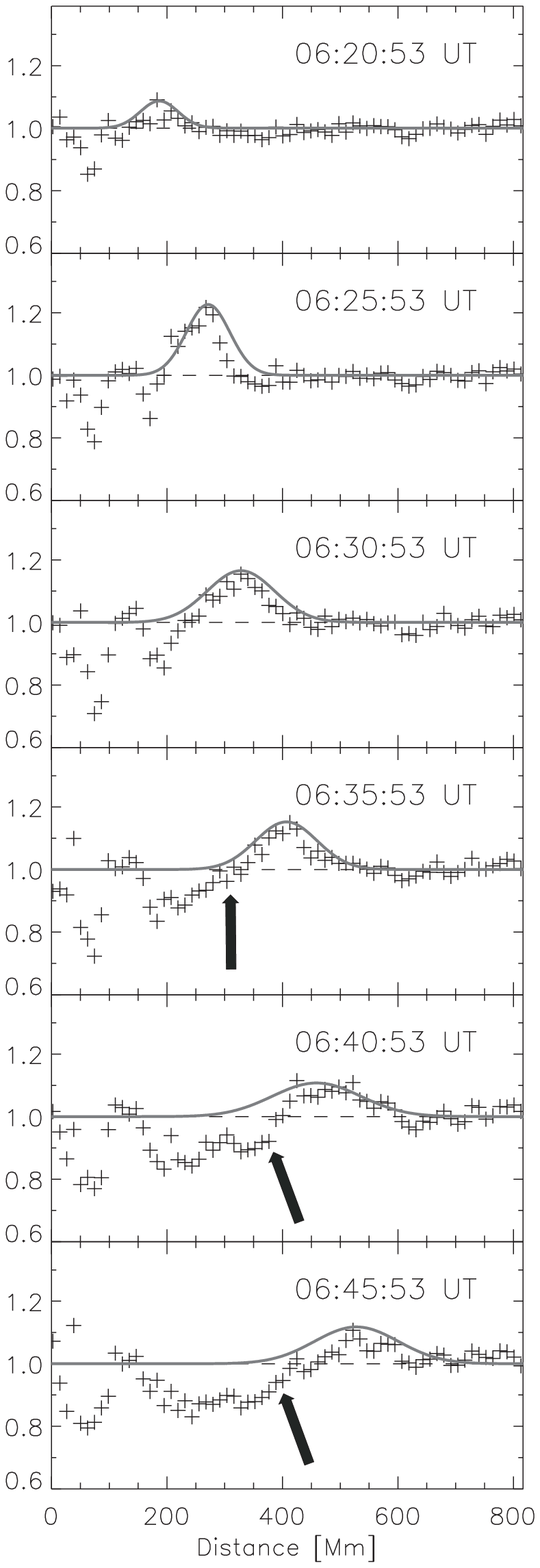}
    \caption[.]{Perturbation profiles of 2010 April 29 ST-B observations in the analyzed sector ($65\degr\pm22.5\degr$)
    together with the gaussian envelopes
    of the wave pulses (thick grey curves). The black arrows point to the rarefaction region.}
\label{img:all20100428195Bprofile_3_rot0panel10}
\end{figure}

\subsection{Distinct features observable in the perturbation profiles after the wave passage}
The inspection of the derived BR perturbation profiles reveals a
diversity of distinct features after the wave passage, including coronal dimmings, stationary brightenings,
and rarefaction regions. In the following, these are discussed individually for each of the four events in
our sample.

\subsubsection{Event of 2007 May 19}\label{subsubsection:2007}
The profiles of 2007 May 19 show
three distinct phases: the leading
wave pulse (enveloped by gaussian fits), a reversal point (where the
intensity level switches from values greater than 1.0 to below 1.0) and the
trailing rarefaction region.
Additionally, in the rear section of the perturbation profiles we observe a
deep core dimming extending from the initiation site up to
$\approx300$~Mm, with an intensity decrease of $\approx70\%$ lasting
for several hours (Fig.~\ref{img:all20070516195Bprofile_1_rot0panel10}).

The rarefaction region and its propagation is of special interest.
This phenomenon is theoretically predicted to be located behind the
propagating front of a compression wave but observations are barely reported
\citep{cohen09}. From 12:52~UT until 13:02~UT we observe the
formation of a deep coronal dimming behind the wave
pulse. During its early evolution it becomes more pronounced and
reaches its maximal spatial extension at 13:02~UT. After 13:02~UT it
is quasi stationary. However in the frontal part of the dimming,
from 300~Mm up to the trailing part of the wave pulse, a smaller,
propagating dip is clearly visible.
We interpret this as the rarefaction region
following the wave front. Its formation is a combination of two aspects:
1) the driver stops at a distance of $\approx300$~Mm (where the
outermost point of the deep-dimming region is located), and 2) behind the
wave pulse the wave forms a region where the
density and pressure fall below the equilibrium level.

This event was associated with a metric type II burst, indicating shock formation
in the corona. We use the observations of the type II burst to obtain an alternative estimate of the shock Mach number. Type II bursts usually show the fundamental and harmonic
emission band, both frequently being split in two parallel lanes,
so-called band-split \citep{nelson85}. The interpretation of the
band split as the plasma emission from the upstream and downstream
shock regions was affirmed by \citet{vrsnak01}. Thus, the band-split
can be used to obtain an estimation on the density jump at the shock
front and thus the Mach number of the ambient plasma
\citep{vrsnak02}.

\begin{figure}
    \centering%
    \includegraphics[width=0.45\textwidth,keepaspectratio=true]{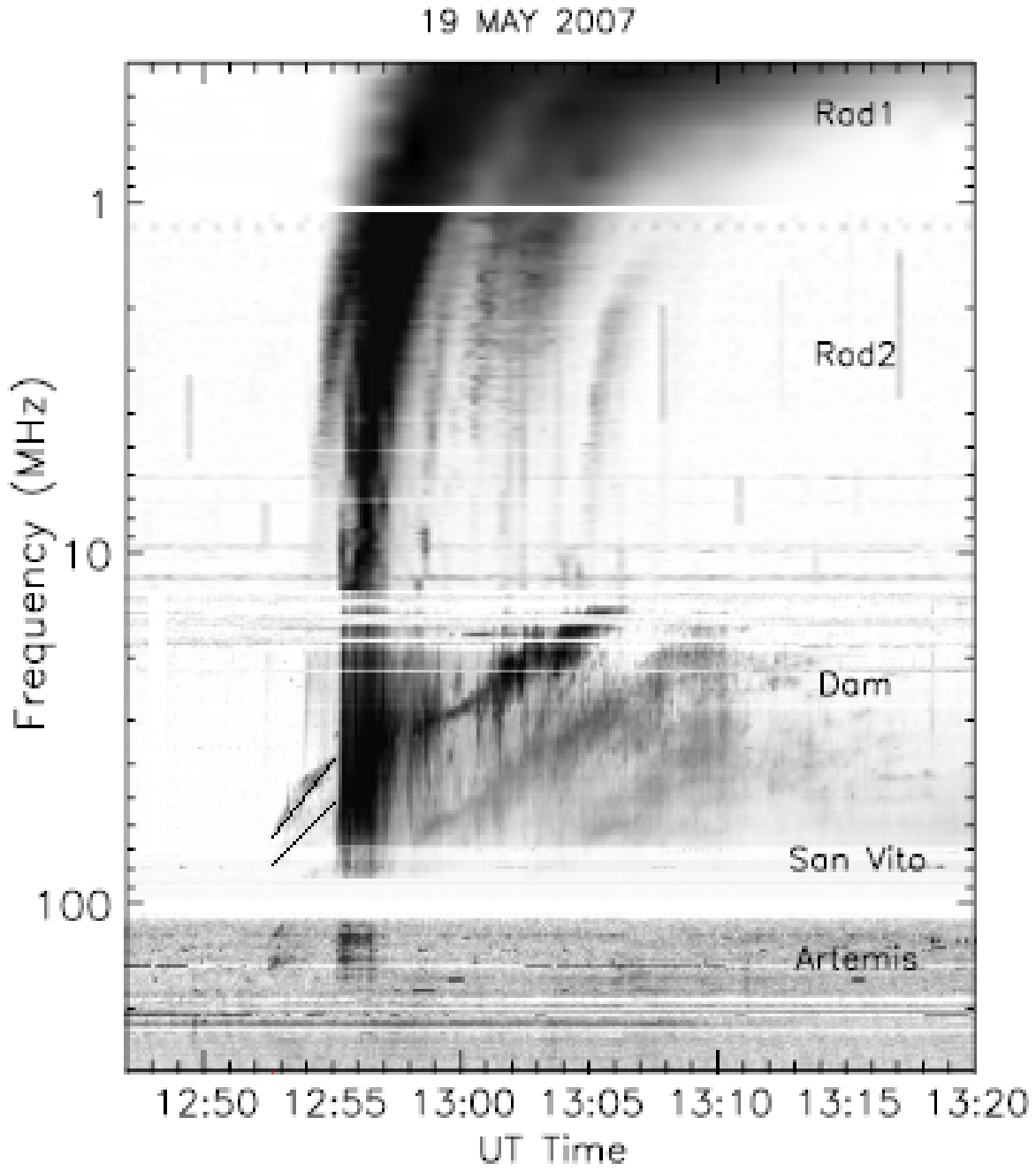}
    \caption[.]{A composite dynamic radio spectrum in the frequency range 0.4\,-–\,300 MHz (adapted from \citet{kerdraon10}).
    Type II and type III bursts are clearly visible. The two branches of the bandsplit of the fundamental emission band
are marked by black lanes.} \label{img:spectrum}
\end{figure}

The composite dynamic
radio spectrum (Fig.~\ref{img:spectrum}) shows complex and intense
radio emission, consisting of a group of type III bursts and a type
II burst. For a detailed dynamic spectrum analysis covering the
frequency range 0.4\,--\,300 MHz, we refer to Fig.~4 in
\citet{kerdraon10}. In the present study we focus on the type II
burst that started around 12:51:30 UT at a frequency of 160 MHz. In
the period 12:51:30~UT to 12:54:00~UT we were able to recognize a
band-split pattern (indicated in Fig.~\ref{img:spectrum} as two
black lines), with a relative bandwidth of $BDW=\Delta f/f\approx
0.19$\,--\,0.22. The relative
bandwidth $BDW$ is determined by the density jump at the coronal
shock front $X_c=\rho_{2c}/\rho_{1c}$, where $\rho_{1c}$ and
$\rho_{2c}$ are the densities upstream and downstream of the shock
front \citep[for details see][]{vrsnak01}. Since BDW$\equiv (f_2-f_1)/f_1=\sqrt{(\rho_{2c}/\rho_{1c})}-1$,
we find $X_c=1.41\,-1.48$. Using a 5-fold Saito coronal density model \citep{saito70}, we find from the derived values of the
density jump $X_c$ a Mach number of 1.32\,--\,1.40.

From the peak amplitude of the EUVI
perturbation profile we derived
$M_{ms}\approx1.21$. A possible explanation for the smaller estimate is that, due to the width of 45$\degr$ over which the perturbation profiles were derived, eventual higher values are averaged out (see
Figure~\ref{img:all20070516195Bprofile_1_rot0panel10}). The curve
displayed by small black crosses shows the overall 45$\degr$ sector
we used for our analysis, while the grey line profile is displaying a
small 2$\degr$ sub-sector in the central part of the overall
45$\degr$ sector. By shrinking the sector to a narrower width around
the most intense propagation direction, we obtain a considerably higher
peak due to a lower signal averaging, and the Mach number is
increased to values of at least $M_{ms}=1.3$. We note that the sector narrowing and the subsequent increased Mach number
implies that the extracted values are (in fact always) underestimated and thus a lower limit for the Mach number estimate. Comparing the results of both
approaches we find the Mach number of the ambient plasma to be
$M_{ms}=1.2$\,--\,1.4, a reasonable range of values for a coronal shock
wave \citep{vrsnak01}. Since EUV waves and type~II bursts are generated at different heights in the solar atmosphere
the Mach numbers are not necessarily the same. In particular, type II bursts are point-like sources formed where the disturbance is most intense, corresponding to higher Mach numbers.
Shocks having Mach numbers in the range of 1.2--1.4 as determined for the events under study are usually considered as weak shocks.

\subsubsection{Event of 2010 January 17}
\label{subsubsection:2010a} The perturbation profiles of 2010
January 17 show persistent stationary brightenings
(Figures~\ref{img:all20100117195Bprofile_2_rot0panel10}). They
appear after the wave front passage and are only slowly fading
away \citep{delanee99, attrill07, delannee07}. The EUV wave starts
at 03:56~UT as a relatively small wave pulse that is intensifying
until 04:01~UT. In the subsequent panels of
Figure~\ref{img:all20100117195Bprofile_2_rot0panel10}, after the
wave has intensified up to its maximum amplitude of 1.6 at
04:01~UT, a strong and long-lasting, quasi-stationary bright
feature becomes clearly visible. We find first evidence for this
stationary parts at 04:06~UT at a distance of 100--300~Mm from the
source location, i.e. at the distance of the first wave front
appearance (see
Figure~\ref{img:all20100117195Bprofile_2_rot0panel10}). It is well
pronounced and exceeds an amplitude level of 1.25, lasting for at
least 30~min after the wave passage. Compared to the time scale of
the propagating disturbance it is only slowly fading away. This
evolution can also be seen in the BR images in
Figure~\ref{img:20100117195Bbaseratio_rot0_overview}, where we
overplotted the position of the stationary brightenings by grey
lines.

\subsubsection{Event of 2009 February 13}
The perturbation profiles of the quadrature event of 2009 February
13 are shown in
Figure~\ref{img:all20090213195Aprofile_1_rot0panel10}. The evolution
of the most enhanced direction of 45$\pm22.5\degr$ is displayed in
the perturbation profiles by crosses. The propagating wave pulse can be clearly identified.
The grey line profile represents a different sector with its main direction to be
270$\pm22.5\degr$. It is characterized by an intensity
enhancement which can be interpreted as a stationary brightening,
showing a similar evolution like the one of 2010 January 17. The
maximum intensity of these stationary brightenings is even more
pronounced than that of the 2010 January 17 event and exceeds a
level of 1.35 at 05:55~UT at 100--300~Mm from the source location
(see Figure~\ref{img:all20090213195Aprofile_1_rot0panel10}). It is
again induced by the wave front passage and only slowly fading away. The propagating wave
pulse ahead of the stationary brightening is not as pronounced as for
the main direction but still detectable although it is considerably smaller than the stationary brightening.
In both directions the coronal dimming is clearly
visible. Its location is restricted to the area behind the
stationary brightening at a distance up to $\approx100$~Mm.

\subsubsection{Event of 2010 April 29}
\label{subsubsection:2010b} The wave event of 2010 April 29 is the
fourth and strongest of four homologous waves that occurred within
a time range of 8 hours \citep{kienreich11}. The profile evolution
of the event is displayed in
Figure~\ref{img:all20100428195Bprofile_3_rot0panel10}. We can
identify three different parts in the rear section of this
perturbation profile: a dimming region, a stationary brightening
and a rarefaction region, all visible after the waves' passage.
The dimming region is prominent from the first image at 06:20~UT
until the last profile shown at 06:40~UT, occurring at 0--100~Mm
from the source location. Right in front of it an intensity
enhancement is present. A definite identification with a
stationary brightening is difficult due to its relatively weak
appearance compared to the background level of 1.0. This
brightening lasts for at least 40~min after the passage and is
only slowly fading away. In front of this intensity enhancement a
dip in the profile is propagating away from the initiation center,
which we interpret as the rarefaction region, similar to the event
of 2007 May 19.

\section{Discussion and Conclusions}\label{sec:discussion}
\begin{enumerate}
  \item We analyzed four well pronounced EUV wave events observed by the
STEREO EUVI telescopes in order to compare two different
kinematical analysis techniques, the generally used visual
tracking method and the semi-automated perturbation profiles. The
differences in the determined positions of the leading edge of the
wave fronts using both methods are maximal 40~Mm. Thus, we
conclude that the perturbation profiles are a suitable method to
analyze large-scale waves. The big advantage of the perturbation
profile method is the higher degree of automatization and
reproductiveness due to the objective measurements of the wave
location. Nevertheless, there are some restrictions. In the later
evolution phase, the wave fronts become more irregular showing
changes in shapes and propagation direction. These effects are not
considered in the profile method. Thus, due to the lower amplitude
and the smearing of irregular fronts the wave profile is no longer
distinct against the background. These properties lead to a
systematic underestimation of the distance of the wave compared to
the visual tracking method. Additionally, the diffuse bright
fronts are easier to identify in RR images due to the fact that
they have a higher contrast than BR images.

\item The determined propagation velocity derived from both methods match within the error limits.
For the four wave events under study, we obtain mean velocity values in the
range of $220-350$~km\,s$^{-1}$, i.e. within the velocity range
for fast magnetosonic waves during quiet Sun conditions
\citep[e.g.][]{mann99}. Three of the four kinematical curves show
only small deceleration values of $-6$ up to $-13$~m~s$^{-2}$,
which lie within the error range.
For the fastest event of 2007 May 19, the obtained deceleration of
$-193$~m~s$^{-2}$ is significant.

\citet{long11} has analyzed the kinematics of the May 19, 2007 and February 13, 2009 events
with a similar perturbation profile method.
They find for the May 19, 2007 event the following values: start velocity $v_0$=447$\pm$87~km~s$^{-1}$, acceleration $a$=--256$\pm$134~m~s$^{-2}$, a broadening of the pulse width from 50 to 200~Mm and a decrease of the peak amplitude intensity from 60 to 15\% of the background values. Our findings are $v_0$=429$\pm$58~km~s$^{-1}$, $a$=--85$\pm$60~m~s$^{-2}$, and similar results for the broadening of the pulse and their intensity evolution. For the second event of February 13, 2009, \citet{long11} find $v_0$=274$\pm$53~km~s$^{-1}$, $a$=--49$\pm$34~m~s$^{-2}$ while our results are $v_0$=261$\pm$72~km~s$^{-1}$, $a$=--24$\pm$30~m~s$^{-2}$. The profile broadening and the intensity evolution is again similar. For the February 13, 2009 event the outcomes of both studies are consistent within the error limits. The results of May 19, 2007 show differences in the acceleration values which can be explained by taking into account that the method used by \citet{long11} and ours is not identical. 1) While they are extracting the position of maximum intensity we are using the leading edge of the wave fronts for our analysis. The difference in the acceleration values is thus due to this usage of different features for the position measurements. The maximum intensity peaks are propagating at a slower speed compared to the leading edges, an effect that is expected for a feature that is broadening during its evolution. Due to that the deceleration values are smaller for the leading edge measurements while the starting velocity values are not affected. 2) The overall sector width used in \citet{long11} is varying from event to event while we are using a constant angular width of 45$\degr$ in each event. 3) The main direction of the calculated sector as well as the the initiation centers are not exactly the same (the uncertainty is about 20~Mm).

\item Considering the events under study as low-amplitude MHD fast-mode
waves the Mach numbers and wave velocity values are proportional
to each other, $M_{ms}=v/v_{ms}$. Thus, from the derived Mach
number evolution (from the maximum value down to $M_{ms}\approx
1$) we expect a decrease of the propagation velocity by $\Delta v
\approx100$ km\,s$^{-1}$ (2007 May 19), $\approx35$ km\,s$^{-1}$
(2009 February 13), $\approx60$ km\,s$^{-1}$ (2010 January 17) and
$\approx30$ km\,s$^{-1}$ (2010 April 29). The velocity changes we
derive from the least-square quadratic fits reveal $\approx 200$
km\,s$^{-1}$ (2007 May 19), $\approx40$ km\,s$^{-1}$ (2009
February 13), $\approx20$ km\,s$^{-1}$ (2010 January 17) and
$\approx30$ km\,s$^{-1}$ (2010 April 29). For the 2009 February 13
and 2010 April 29 events, the derived velocity changes are in the
same order of $\approx30$ km\,s$^{-1}$ as the error on the
velocity determination. Hence, the weak deceleration is hidden in
the measurement uncertainties. For the 2007 May 19 event the
result of $200$ km\,s$^{-1}$ clearly exceeds the error in velocity
of $\approx70$ km\,s$^{-1}$, thus the deceleration is observable
and significant. According to the Mach number evolution for the
2010 January 17 event, an observable deceleration of the
propagating wave is expected. The determined velocity change of
$\approx60$ km\,s$^{-1}$ exceeds the velocity error of $20$
km\,s$^{-1}$, thus masking of the deceleration by data scatter is
unlikely. Nevertheless, from the kinematical measurements obtained
with the visual tracking and the perturbation profiles we do not
derive a significant deceleration, as it would be expected from
the peak Mach number estimated for this event.

  \item Each
event can be followed over a period of at least 30 minutes, during
which we first observe an intensification of the wave, which is
followed by a steady decrease and broadening of the wave pulses as
extracted from the perturbation profiles. From this we derived
useful information on the characteristics of the wave pulses, like
the peak amplitude, the width, and the integral of the wave pulse.
The evolution of the pulse width shows a clear broadening over
time by a factor of 2--3, similar to the broadening of the FWHM
values. These values are typical for large-scale waves
\citep{warmuth04b, wills06, warmuth10, veronig10}. The integral
below the perturbation profile basically remains constant (2007
May 19 and 2010 April 29) or decreases by a factor of 3\,--\,4
(2009 February 13 and 2010 January 17). We want to stress that the
combination of profile broadening and amplitude decay leading to a
constant integral below the disturbance profile is consistent with
the characteristics of a freely propagating wave \citep{landau87}.

\item The peak magnetosonic Mach numbers derived for the events under study are
in the range of $M_{\rm{ms}}=1.08$\,--\,1.21, indicative of the
evolution of weak coronal shocks. The magnetosonic Mach number for
2007 May 19 was determined via two different approaches leading to
similar results in a range of $M_{\rm{ms}}=1.2$\,--\,1.4. Thus, the
usage of the perturbation maximum amplitude for the determination of
the magnetosonic Mach number seems reasonable. \citet{grechnev11}
discussed the problem of projection effects in measurements of EUV
wave. Since we do not observe the perturbation from a view-point
that is located directly above the propagating disturbance but from
an inclined position, the intensity values are always
underestimated. This is probably also the reason why we do
observe Gaussian wave pulse profiles instead of a sharp edge,
expected in the case of a shocked disturbance.

\item We observe three different distinct features in
the perturbation profiles occurring after the wave passage.
In
three of four events deep coronal dimmings are prominent at
small distances from the eruption center.
Only for the 2010, January 17 event it is hardly present in
the perturbation profiles of the used sector. However, an inspection of
Figure~\ref{img:20100117195Bbaseratio_rot0_overview} reveals the
existence of coronal dimmings in the near vicinity of the initiation
site most prominent to the western sectors close to the limb and
also off-limb.

The second distinct feature in the wake of three of the four the EUV
waves are stationary brightenings. These stationary parts appear
after the waves' passage at distances of 100--300~Mm and fade only
slowly away. Due to the fact that the the most intense wave profile
always forms one frame before the stationary brightening appears,
these parts seem to be an obstacle for the wave front.
These stationary brightenings are not circumferential around the
eruption center but confined to specific sectors, which is best
observed for the event of 2009 February 13. \citet{cohen09} did numerical simulations on
this event and showed that stationary brightenings are developing
at the outer edges of the core coronal dimmings. Observations by \citet{delanee99, delanee00} and \citet{delannee08}
as well as simulations by \citet{chen05b}
indicate that these stationary brightenings appear at the
locations where the connectivity of the magnetic field lines changes.
Since these stationary brightenings are located in front of the deep core dimming regions (interpreted as the footprints of the expanding CME), are restricted in their expansion (their maximal distance is given by the first wave fronts location) and appear only in confined sectors after the waves' passage, they can be interpreted as a signature of the CME expanding flanks.

The third distinct feature are rarefaction regions observable in
two of four wave events, on 2007 May 19 and 2010 April 29. The
rarefaction region behind the wave pulse is a prospective feature
that develops trailing a large-amplitude perturbation
\citep{landau87}. Indeed, simulations by \citet{cohen09} revealed
regions, which are dim in intensity and propagate after the bright
wave front.
\end{enumerate}

The EUV wave velocities are in the range of fast magnetosonic
waves in the quiet solar corona and show either a decelerating
characteristics or constant velocity (within the measurement
uncertainties) during propagation. The perturbation profiles
reveal intensification, broadening and decay of the propagating
wave pulse, while the integral evolution is either constant, or
decreasing. The small Mach numbers are indicative of a weak
coronal shock. Stationary brightenings formed in confined sectors
at the outer edge of the core coronal dimming can be interpreted
as signatures of the expanding CME flanks. The first observable
wave fronts are always located right in front of them. Thus, we
conclude that the EUV wave events can be interpreted as freely
propagating, weak shock fast-mode MHD waves initiated by the CME
lateral expansion.

\acknowledgments

We thank the STEREO/SECCHI teams for their open data policy. N.M.,
I.W.K., and A.M.V. acknowledge the Austrian Science Fond (FWF):
P20867-N16. The European Community's Seventh Framework Programme
(FP7/2007-2013) under grant agreement no. 218816 (SOTERIA) is
acknowledged by B.V. and M.T.

\bibliographystyle{aa}


\end{document}